\documentclass[useAMS,usenatbib]{mn2e}

\usepackage{epsfig}
\usepackage{amsmath}
\usepackage{amssymb}
\usepackage{natbib}
\usepackage{threeparttable} 
\usepackage[hyphens]{url}

\usepackage{rotating}
\usepackage{graphicx}
\usepackage{epstopdf}

\usepackage{color}

\usepackage{booktabs,array}

\newcommand{\ax}{AX J1754.2--2754}
\newcommand{\rxs}{1RXS J171824.2--402934}
\newcommand{\rxh}{1RXH J173523.7--354013}
\newcommand{\xte}{XTE J1719--291}

\newcommand{\ascasource}{AX J1538.3--5541}
\newcommand{\xmmsource}{XMMU J174716.1--281048}

\newcommand{\chan}{\textit{Chandra}}
\newcommand{\swift}{\textit{Swift}}
\newcommand{\xmm}{\textit{XMM-Newton}}

\newcommand{\rosat}{\textit{ROSAT}}
\newcommand{\beppo}{\textit{BeppoSAX}}
\newcommand{\asca}{\textit{ASCA}}
\newcommand{\inte}{\textit{INTEGRAL}}

\newcommand{\lum}{\mathrm{erg~s}^{-1}}
\newcommand{\flux}{\mathrm{erg~cm}^{-2}~\mathrm{s}^{-1}}
\newcommand{\cnts}{\mathrm{counts~s}^{-1}}
\newcommand{\nh}{\mathrm{cm}^{-2}}

\newcommand{\qui}{\mathrm{\chi}^{2}_{\nu}}

\title[Very-faint persistent X-ray binaries]{The X-ray spectral properties of very-faint persistent neutron star X-ray binaries}
\author[M. Armas Padilla et al.]
{M. Armas Padilla$^{1}$\thanks{e-mail: m.armaspadilla@uva.nl},
N. Degenaar$^{2}$\thanks{Hubble fellow}, 
R. Wijnands$^{1}$\\ 
$^{1}$Astronomical Institute ``Anton Pannekoek", 
University of Amsterdam, 
Postbus 94249, 1090 GE Amsterdam, The Netherlands\\
$^{2}$Department of Astronomy, 
University of Michigan, 
500 Church Street, Ann Arbor, MI 48109, USA\\
}

\begin{document}

\date{DRAFT VERSION}

\pagerange{\pageref{firstpage}--\pageref{lastpage}} \pubyear{0000}

\maketitle

\label{firstpage}

\begin{abstract} \ax, \rxs\ and \rxh\ are three persistent neutron
star low-mass X-ray binaries that display a 2--10 keV accretion
luminosity $L_{\mathrm{X}}$ of only $(1-10) \times 10^{34}~\lum$ (i.e., only
$\simeq$0.005-0.05\% of the Eddington limit). The phenomenology of
accreting neutron stars which accrete at such low accretion rates is
not yet well known and the reason why they have such low
accretion rates is also not clear. Therefore, we have obtained \xmm\
data of these three sources and here we report our analysis of the
high-quality X-ray spectra we have obtained for them. We find that \ax\ has
$L_{\mathrm{X}}\sim 10^{35}~\lum$, while the other two have X-ray
luminosities about an order of magnitude lower. However, all sources have a similar, relatively soft, spectrum with a photon index of 2.3-2.5, when the spectrum is fitted with an absorbed power--law model. This model
fits the data of \ax\ adequately, but it cannot fit the data obtained
for \rxs\ and \rxh. For those sources a clear soft thermal component
is needed to fit their spectra. This soft component contributes 40\%-50\% to the 0.5--10~keV flux of the sources. The presence of this soft component might be the reason why the spectra of these two sources are soft. When including this additional spectral component, the power-law photon indices are significantly lower. It can be excluded that a similar component 
with similar contributions to the 2--10~keV X-ray flux is
present for \ax, indicating that the soft spectrum of this
source is mostly due to the fact that the power-law component itself
is not hard. We note that we cannot excluded that weaker soft component is present in the spectrum of this source which only contributes up to $\sim25\%$ to the 0.5--10~keV X-ray flux. We discuss our results in the context of what is known of
accreting neutron stars at very low accretion rate.  
\end{abstract}


%


\section{Introduction}\label{sec:intro}

Low mass X-ray binaries (LMXBs) are systems in which a neutron star or
black hole accretes material from a Roche-lob filling companion star
that typically is less massive than the accretor. Transient X-ray
binaries alternate long-lived quiescent periods (typically years to
decades) during which the systems are extremely faint in the X-rays
($< 10^{33}\lum$; during which no or hardly any accretion occurs) with short-lived (weeks to months) accretion
outburst episodes during which the X-ray luminosity increases
several orders of magnitude. It is thought that those outbursts are
powered by a similar large increase in the mass accretion rate onto
the central compact object.  On the other hand, the persistent systems
are continuously accreting and therefore their X-ray
luminosities are always order of magnitudes higher than the X-ray luminosities in quiescence. Although large fluctuations in
their X-ray brightness are 
commonly observed (sometimes fluctuations of more than a factor of 10), they do not become as faint as the X-ray transients in quiescence. 

Very faint X-ray binaries (VFXBs) are those systems that show maximum 2--10~keV 
X-ray luminosities of only $\sim10^{34-36}\lum$
\citep{Wijnands2006}. These sub-luminous sources are interesting
because they probe a relatively unexplored mass-accretion regime and
can therefore provide valuable input for accretion physics
\citep[e.g.,][]{armaspadilla2011,armaspadilla2013}, binary evolution
models \citep[e.g.,][]{king2006,degenaar2009,degenaar2010,Maccarone2013} and the
theory of nuclear burning on the surface of accreting neutron stars
\citep[e.g.,][]{cooper2007,peng2007,degenaar2010_rxh}.  

At the time of writing, only a handful of VFXBs persistently accreting at such
very low accretion rates are known
\citep[e.g.,][]{zand2005,zand2007,degenaar2010_rxh,degenaar2012_asca,degenaar2012_igrburster}. The very faint persistent systems for which the accretor has conclusively been identified harbour a neutron star because they
exhibited a type-I burst. However, the fact that we do not know any
black hole system is very likely a selection effect because without
observing type-I bursts or pulsations, the nature of the accretor is very difficult
to identify. The reason why those sources exhibit such faint accretion luminosities remains unclear.
It has been proposed that persistent VFXBs could be ultra compact X-ray binaries \citep[UCXBs;][]{King2000}. These are systems that are
thought to harbor an hydrogen-poor companion star in a very small
orbit  around the accretor with an orbital period $P \lesssim 80-90$~min
\citep[e.g.,][]{nelson1986,nelemans2010}. However, although a very short orbital period could justify the low luminosity for some sources, it can not explain all persistent VFXBs \citep[][see section 1.3]{degenaar2010_rxh}.
\\

In this work we report on the analysis of high-quality
\xmm\ spectra of three persistent sub-luminous neutron star LMXBs: \ax, \rxs\ and \rxh.

\subsection{\ax}\label{subsec:ax}

\ax\ is a faint X-ray source that was discovered with \asca\ in 1999
\citep[][]{sakano2002}. The nature of the source remained unknown
until \inte\ detected a type-I X-ray burst in 2005, which
unambiguously identified it as an accreting neutron star in a LMXB
\citep[][]{chelovekov2007}. Assuming that the peak X-ray burst was
equal to or lower than the Eddington luminosity, the source distance
was estimated to be $D\la6.6$~kpc for a hydrogen atmosphere of the neutron star, and $D\la9.2$~kpc for a helium atmosphere.

The source has been detected at a 2--10 keV luminosity of
$L_{\mathrm{X}}\simeq10^{35}~\lum$ every time an X-ray satellite
pointed at it
\citep[][]{delsanto2007_ax,krivonos2007,bassa2008,jonker2008,degenaar2012_asca,maccarone2012},
except for a brief episode in 2008, when \chan\ did not detect the
source with an upper limit of $L_{\mathrm{X}}\simeq5 \times
10^{32}~\lum$ \citep[][]{bassa2008}. The non-detection might have been
caused by an eclipse, although this scenario is not very likely
\citep[][]{jonker2008}. Instead it is more likely that the source was
inactive for a brief period, which must have been rather short
\citep[$<1$~yr;][]{degenaar2012_asca}. This inactive period would
indicate that the source is a transient but its unusual short inactive
phase compared to the very long active phase makes this system quite an
oddball among the transients. Since the source is active for such an extended period of time, we expect that the accretion flow geometry in this source is likely to be very similar to that present in the persistent sources and therefore we treat the source as a persistent source in the remainder of the paper.

The source has a low absolute optical magnitude,
which has let to the suggestion that it might be an ultra-compact
X-ray binary \citep[][]{bassa2008}. However, no orbital period
is known for this source so its classification as an UCXB is still
uncertain.

Analysis of a series of \swift\ observations suggests that the X-ray
spectrum of \ax\ is strongly absorbed
(with a column density $N_{\mathrm{H}}\simeq2.1\times10^{22}~\nh$) and relatively soft
\citep[with a photon index $\Gamma \simeq2.7-3.6$;][]{DelSanto2008,degenaar2012_asca}. The \swift\ spectral
data could be adequately fitted with a single power--law model, but
the data quality was such that a multiple component spectrum could not
be excluded.  


\subsection{\rxs}\label{subsec:rxs}
\rxs\ was discovered with \rosat\ in 1990 and identified as a neutron star LMXB when a type-I X-ray burst was detected with \beppo\ in 1997 \citep[][]{kaptein2000}. The X-ray burst showed evidence for photospheric radius expansion. Assuming Eddington-limited emission during the burst peak places the source at a upper limit distance of $D\la6.5$~kpc for a H-rich photosphere, or $D\la9$~kpc if a helium atmosphere is assumed \citep[][]{zand2005}. The source is persistently detected at a 2--10 keV persistent luminosity of $L_{\mathrm{X}} \simeq (5-10) \times 10^{34}~\lum$ \citep[][]{zand2005,zand2009}.  A \chan\ observation showed that the X-ray spectrum of the source can be described by a power--law model with $N_{\mathrm{H}} \simeq 1 \times 10^{22}~\nh$ and $\Gamma \simeq 2.1-2.4$ \citep[][]{zand2005,zand2009}.

\subsection{\rxh}\label{subsec:rxh}
\rxh\ is a faint X-ray source that was discovered with \rosat\ in 1990. It was identified as a neutron star LMXB when \swift\ detected a type-I X-ray burst  in 2008 \citep[][]{israel2008,degenaar2010_rxh}. From the analysis of the X-ray burst a source distance of $D\la9.5$~kpc is suggested, assuming that the peak
flux was equal to or lower than the Eddington luminosity and a helium atmosphere for the neutron star. When a H-rich atmosphere is assumed, the distance lowers to $D\la6.2$~kpc. Its X-ray spectrum can be described by an absorbed power--law with $N_{\mathrm{H}} \simeq 9 \times 10^{21}~\nh$ and $\Gamma \simeq 2.3$. The source is always detected at a 2--10 keV luminosity of $L_{\mathrm{X}}\simeq10^{35}~\lum$ \citep[][]{degenaar2010_rxh}. 

\rxh\ has an optical/IR counterpart with observed magnitudes of $V\simeq21.2$, $R\simeq18.8$, $J\simeq15.4$, $H\simeq14.3$, and $K\simeq13.8$~mag. Optical spectroscopy revealed a strong H$\alpha$ emission line, which suggests that the neutron star is accreting H-rich material \citep[][]{degenaar2010_rxh} and therefore the system cannot be a UCXB.


\section{Observations and data analysis}\label{sec:obs}

In order to investigate the spectral properties of VFXBs in greater
detail than previous studies, we acquired \xmm\ observations of the
three VFXBs discussed in section \ref{sec:intro} (Table~\ref{tab:obs}
shows a log of the observations). We use the data obtained with the
European Photon Imaging Cameras (EPIC), which consists of two MOS
detectors \citep[][]{turner2001} and one PN camera
\citep[][]{struder2001}. During all three observations the MOS cameras
were operated in imaging mode using the small window for \rxh\ and the
full frame window for \ax\ and \rxs , whereas the PN was set in timing
mode for the three observations. The timing mode was used to allow
searches for the expected millisecond X-ray pulsations in these
sources arising from a possible fast rotating neutron star in them
\citep[e.g.][]{Wijnands2008}. However no pulsations were found (see
\citealt{Patruno2010} for the results on \rxs ; a full description will be
published elsewhere). Given the faintness of our sources and the calibration uncertainties in the soft
part of the spectrum in the PN data in timing mode \citep[see the XMM-Newton calibration technical note XMM-SOC-CALTN-
0083\footnote{\url{http://xmm.esac.esa.int/}}; see also, e.g., ][]{Done2010}, we have excluded the PN spectra from our analysis.\\

The data analysis and reduction were carried out with the Science
Analysis Software (\textsc{sas}, v. 12.0.1) to obtain calibrated event
lists and scientific products. The observations were affected by
episodes of high background. To eliminate these background flaring
periods we excluded data with count rates larger than
0.35~$\cnts$ at energies $>10$~keV. The count rates of the three observations were lower than 0.55~$\cnts$ (see table \ref{tab:obs}), while the pile starts to be an issue at counts rates  above $\sim0.7$~$\cnts$ (see the \xmm\ users'
handbook\footnote{\url{http://xmm.esac.esa.int/external/xmm_user_support/documentation/uhb_2.1/XMM_UHB.html}}). Therefore, our sources are not bright enough for pile-up to become considerable and there was not need to correct for that. We
extracted the MOS source events using a circular region with a radius of $\sim 950$,
$\sim 1000$ and $\sim 700$ pixels for \ax, \rxs\ and \rxh,
respectively. The background events were extracted from circles with
the same sizes and were located on nearby source-free regions. We
generated the light curves and the spectra, as well as the associated
response matrix files (RMFs) and ancillares response files (ARFs)
using the standard analysis
threads\footnote{\url{http://xmm.esac.esa.int/sas/current/documentation/threads/}}. The
spectral data were grouped to contain a minimum of 25 photons per bin
and rebinned to oversample the \textit{FWHM} of the energy resolution
by a factor of three.\\

For each observation, we fitted simultaneously the 0.5--10~keV spectra
of the two MOS cameras using \textsc{XSpec}
\citep[v. 12.8;][]{xspec}. The model parameters were tied between the
two detectors. We added a constant factor (\textsc{constant}) to the
spectral models with a value fixed to one for MOS1 camera but we allowed
it to vary freely for MOS2 in order to account for
cross--calibrations uncertainties. We included the photoelectric
absorption component (\textsc{phabs}) to account for the interstellar
absorption assuming the cross--sections of \citet{verner1996} and the
abundances of \citet{Wilms2000}. \\

\begin{table*}
\caption{Target list and \xmm\ observation log.}
\begin{threeparttable}
\begin{tabular}{l c c c c c}
\hline \hline
Source  & Observation ID & Date & Exposure time & Net exposure time$^{a}$ & Net count rate\\
&  & (yyyy-mm-dd) & (ks) & (ks) & (counts s$^{-1}$) \\
\hline
\ax\ & 0651450201 & 2011-03-17 & 47.5 & 42 & 0.54\\ 
\rxs\  & 0605160101 & 2010-03-14  & 76.8 & 70 & 0.25\\ 
\rxh\  & 0606200101 & 2010-03-20 & 40.0 & 28 & 0.17\\ 
\hline
\end{tabular}
\label{tab:obs}
\begin{tablenotes}
\item[a]{Net expesure time after removing episodes of background flaring}
\end{tablenotes}
\end{threeparttable}
\end{table*}


\section{Results}\label{sec:results}

We used six different models to fit our spectra. First we used a
simple power--law (\textsc{powerlaw}) affected by photoelectric
absorption, and then we added a soft component, using either a
blackbody (\textsc{bbodyrad}) or an accretion disk model consisting of
multiple blackbody components (\textsc{diskbb},
\citealt{Makishima1986}). In order to investigate the effects of our choice of model for the hard component on our results, we repeated the fits replacing the \textsc{powerlaw} with a thermally comptonized continuum model \citep[\textsc{Nthcomp;}][]{Zdziarski1996,.Zycki1999}. In the fits with the two component models, we tied the seed photon temperature (low energy rollover; $kT_{\mathrm{bb}}$ ) to the blackbody temperature when using \textsc{bbodyrad} model, and to the temperature at inner disk radius when using the \textsc{diskbb} model. The obtained photon index and temperatures are, within the uncertainties, compatible with the values obtained when using the \textsc{powerlaw}, although we note that for \ax\ the fit using the \textsc{Nthcomp} is unstable inhibiting us to correctly determine the errors. For simplicity, in the main body of the paper we only discuss the \textsc{powerlaw} results, which are
summarized in Table~\ref{tab:spec}. The uncertainties on the spectral
parameters are at 90\% confidence level and the flux errors have been
calculated following the procedure presented by
\citet{Wijnands2004}. All three neutron stars exhibited very energetic thermonuclear X-ray bursts that were likely caused by the ignition of a layer of pure He. We therefore adopt the distances that were inferred by assuming pure He bursts when calculating luminosities and blackbody/diskbb emitting radii. These are 9.2~kpc for \ax\ \citep{chelovekov2007}, 9~kpc for \rxs\ \citep{zand2005}, and 9.5~kpc for \rxh\ \citep{degenaar2010_rxh}. \\

\subsection{\ax}

The spectrum of \ax\ can be well described by a single power--law model
($\qui\sim0.96$ for 279 degrees of freedom [dof]; Figure~\ref{fig:spec}). The $N_{H}$ obtained is $\sim2.9\times10^{22}~\nh$. The X-ray spectra
are quite soft: the obtained photon index is $\Gamma=2.5$, which is
(within the uncertainties) in agreement with the value ($\Gamma=2.7\pm0.2$) obtained
by \citet[][]{degenaar2012_asca} in their analysis of \swift\ data of
the source (however, see \citealt[][]{DelSanto2008} who found a higher value; $\Gamma=3.6\pm0.7$). Adding a soft spectral does not
improve the fit significantly ($\qui\sim0.93$ for 277 dof). When we do add a soft
component to the power--law model, the photon index becomes
$\Gamma\sim2.3$ or $\Gamma\sim2$, and the temperatures obtained are 0.53 or 0.74 keV
for the blackbody and diskbb, respectively. The fractional
contribution of the thermal component to the total unabsorved
0.5--10~keV and 2--10~keV fluxes are almost the same, with a higher
contribution if we use the diskbb ($\sim26\%$) than using the bbodyrad
($\sim8\%$). For all fits the unabsorbed 2--10~keV flux obtained is
$\sim9.5\times10^{-12}~\flux$ which corresponds to a luminosity of
$\sim1\times10^{35}~\lum$ assuming a distance of 9.2~kpc.\\


\subsection{\rxs}
The spectrum of \rxs\ is not well described by a single power--law model ($\qui\sim2.5$ for 277 dof; see the residuals shown in Figure~\ref{fig:spec}) and it is required to add an additional component (for which we use a soft, thermal component) to get an acceptable fit ($\qui\sim1.03$ for 275 dof; Figure~\ref{fig:spec}). The obtained column density is $\sim2\times10^{22}~\nh$ and the photon index $\Gamma\sim1.5$, which is considerably harder than the $\Gamma=2.3$ obtained with the simple power--law model. The temperature of the blackbody (diskbb) component is 0.31~keV (0.39~keV), which contributes 37\% (49\%) in the total 0.5--10~keV unabsorbed flux and only 9\% (11\%) in the total 2--10~keV one. The unabsorbed flux is $\sim4\times10^{-12}~\flux$ (2--10~keV) when we use the bbodyrad component. Assuming a distance of 9~kpc the corresponding luminosity is $\sim3.6\times10^{34}~\lum$. The unabsorbed values are slightly higher when using the diskbb model.\\

\subsection{\rxh}

\begin{figure}$\phantom{!h}$
\begin{center}


\hspace{-0.5cm}
\includegraphics[angle=-90,width=9.5cm]{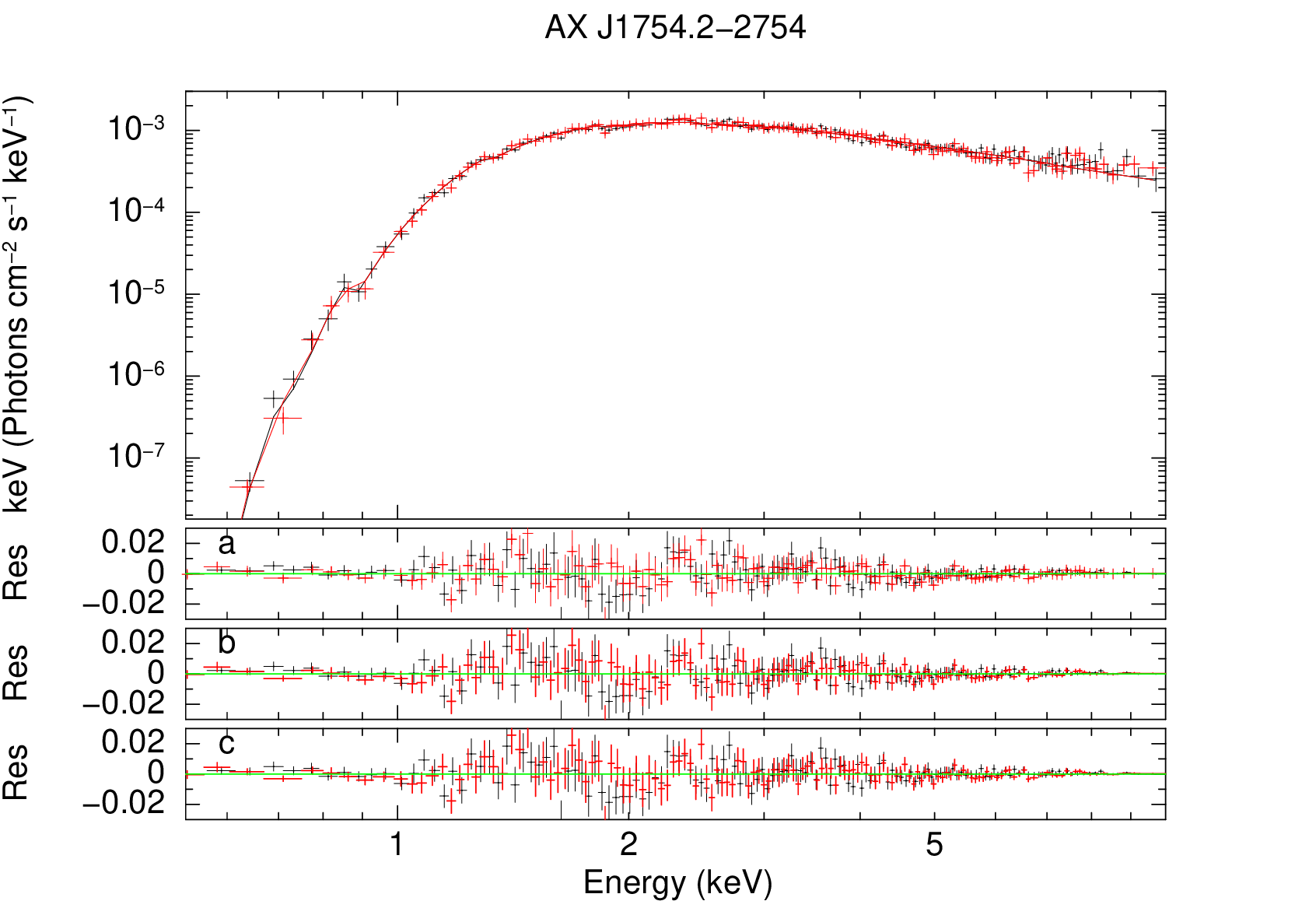}\\
\hspace{0.5cm}
\includegraphics[angle=-90,width=9.5cm]{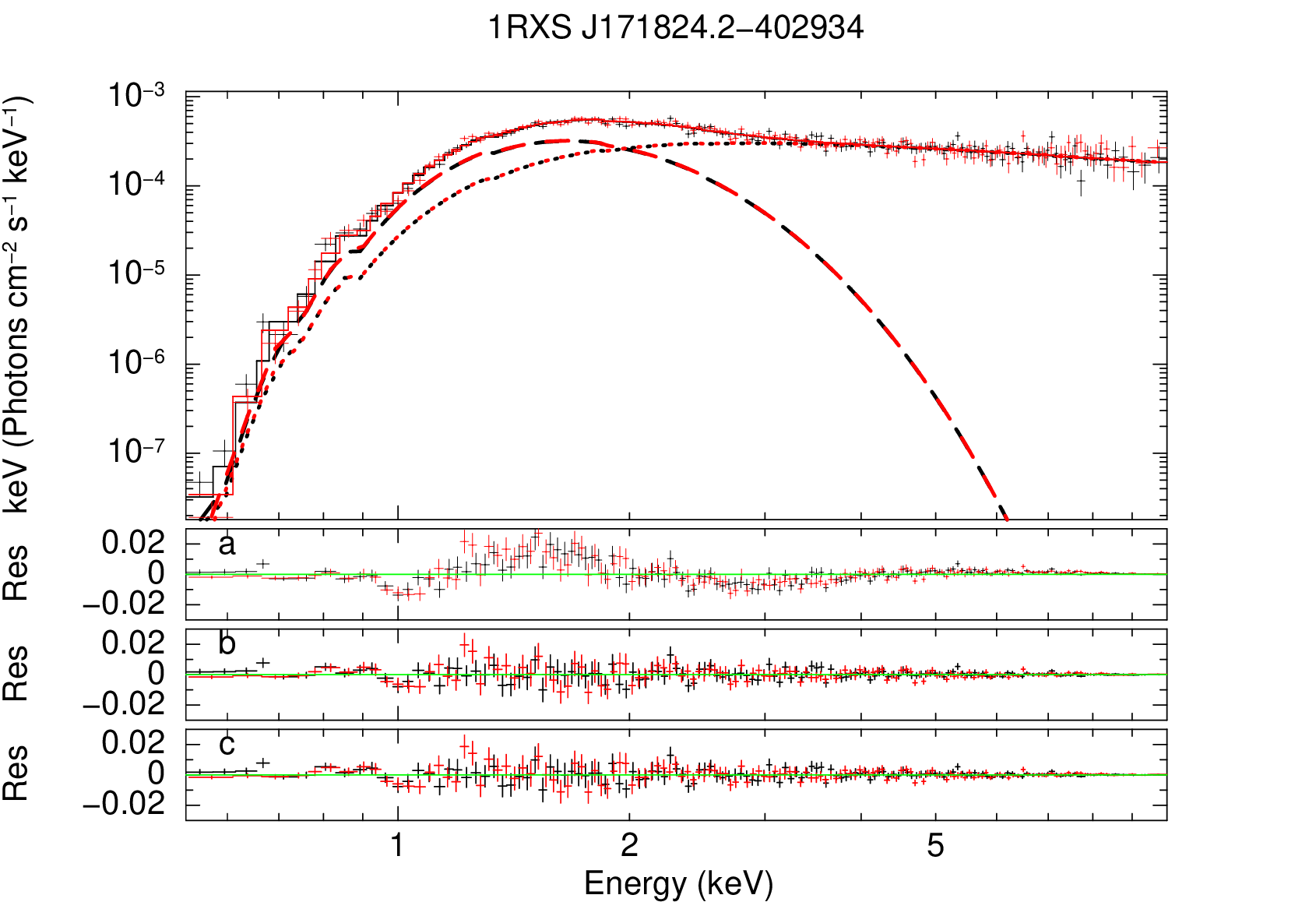}\\
\hspace{0.5cm}
\includegraphics[angle=-90,width=9.5cm]{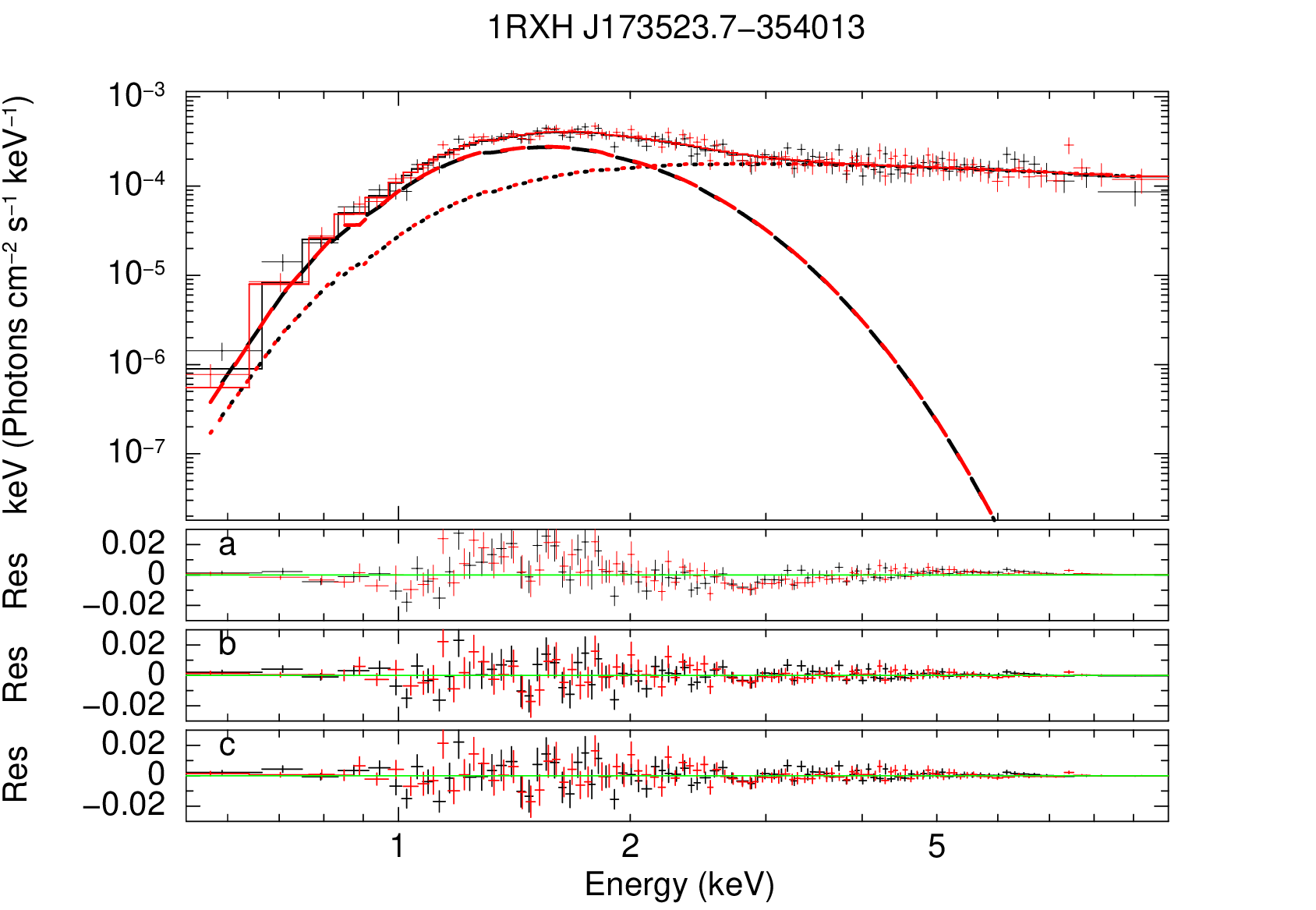}


\caption{ MOS1 (black) and MOS2 (red) spectra. The solid lines represent the best fit for a \textsc{powerlaw} model in the case of \ax\ (top panel), and a combined \textsc{powerlaw} (dashes) and \textsc{bbodyrad} (dotted lines) model for \rxs\ (middle panel) and \rxh\ (bottom panel). In the sub-panels the residuals are plotted  when fitting with a) \textsc{powerlaw}, b) \textsc{powerlaw+bbodyrad} and c) \textsc{powerlaw+diskbb} models}
\label{fig:spec}
\end{center}
\end{figure}

\rxh\ has similar spectral characteristics as \rxs. It also needs a two component model to obtain a good fit ($\qui=0.97$ for 184 dof for a two component model versus $\qui=1.92$ with 186 dof for a single power-law model; Figure~\ref{fig:spec}). The thermal component contributes about half of the total 0.5--10~keV unabsorbed flux while only $\sim10\%$ in the 2--10~keV energy range. The parameters obtained are $N_{H}\sim1.5\times10^{22}~\nh$, a photon index of $\Gamma\sim1.4$ and a temperature of 0.3~keV for the bbodyrad (slightly higher if the diskbb component is used; 0.4~keV). The unabsorbed 2--10~keV flux is $\sim2.3\times10^{-12}~\flux$ which translates to a luminosity of $\sim2.5\times10^{34}~\lum$ for a distance of 9.5~kpc.   \\

%
\begin{table*}
\caption{Results from fitting the spectral data.}
\scriptsize
\begin{sideways}
\begin{threeparttable}
\begin{tabular}{l c c c c c c c c c c c c c}
\hline \hline
Source & $N_{\mathrm{H}}$ & $\Gamma$ & $kT$& $R_{\mathrm{bb/in}}$ & $T_{\mathrm{fr}}$ & $F_{\mathrm{X, abs}}$ & $F_{\mathrm{X,unabs}}$ & $L_{\mathrm{X}}$ & $T_{\mathrm{fr}}$ & $F_{\mathrm{X, abs}}$ & $F_{\mathrm{X,unabs}}$ & $L_{\mathrm{X}}$ & $\chi^2_{\nu}$ (dof) \\
 & ($10^{22}$ $\nh$) & & (keV) &(km) & (\%) &  \multicolumn{2}{c}{($10^{-12}$ $\flux$)}  & ($10^{34}$ $\lum$) &  (\%) &  \multicolumn{2}{c}{($10^{-12}$ $\flux$)}  & ($10^{34}$ $\lum$) & \\ 
 \cmidrule(r){6-9} \cmidrule(r){10-13}
 & &  & &  &   \multicolumn{4}{c}{(0.5--10 keV)}   &  \multicolumn{4}{c}{(2--10 keV)} & \\ 
 
\hline
\multicolumn{14}{c}{phabs*(powerlaw)}\\
\hline
AX~J1754 & $2.93 \pm 0.06$ & $2.51 \pm 0.03$ &  -- & -- & --& $8.81 \pm 0.06$ &  $27.0 \pm 0.8$& $27.4 \pm 0.8$ & -- & $7.70 \pm 0.06$ &  $9.55 \pm 0.02$& $9.67 \pm 0.02$ & 0.96 (279) \\ 
1RXS~J1718 & $1.78 \pm 0.04$ & $2.33 \pm 0.04$ &-- & -- & -- & $3.59 \pm 0.04$ &  $8.0 \pm 0.2$& $7.8 \pm 0.1$ & -- & $2.95 \pm 0.04$ &  $3.33 \pm 0.03$& $3.23 \pm 0.02$ & 2.54 (277) \\ 
1RXH~J1735 & $1.43 \pm 0.07$ & $2.45 \pm 0.07$  & -- & -- &-- &  $2.29 \pm 0.05$ & $5.2 \pm 0.3$& $5.6 \pm 0.3$ & -- & $1.76 \pm 0.05$ & $1.93 \pm 0.05$& $2.08 \pm 0.05$ & 1.92 (186) \\ 
\hline 
\multicolumn{14}{c}{phabs*(powerlaw+bbodyrad)}\\
\hline
AX~J1754 & $2.7 \pm 0.2$ & $2.3 \pm 0.2$ &  $0.53 \pm 0.04$ & $1.4 \pm 1.2$ & 8.4 & $8.90 \pm 0.08$ &  $22 \pm 3$& $22 \pm 3$ & 9 & $7.79 \pm 0.09$ &  $9.4 \pm 0.06$& $9.57 \pm 0.06$ & 0.93 (277) \\ 
1RXS~J1718  & $1.9 \pm 0.1$ & $1.6 \pm 0.1$ &  $0.31 \pm 0.02$ & $ 5.1^{+0.5}_{-0.7}$ & 37.2 & $4.01 \pm 0.05$ &  $7.8 \pm 0.5$& $7.6 \pm 0.2$ & 9.4 & $3.34 \pm 0.05$ &  $3.75 \pm 0.04$& $3.63 \pm 0.02 $ & 1.03 (275) \\ 
1RXH~J1735 & $1.4 \pm 0.1$ & $1.4 \pm 0.2$  & $0.30 \pm 0.02$ &$4.8^{+0.6}_{-1.1}$& 42.2 &  $2.66 \pm 0.07$ & $4.8 \pm 0.5$ & $5.2 \pm 0.5$ & 10 &  $2.10 \pm 0.06$ & $2.29 \pm 0.06$ & $2.48 \pm 0.06$  & 0.97 (184) \\
\hline 
\multicolumn{14}{c}{phabs*(powerlaw+diskbb)}\\
\hline
AX~J1754 & $2.6 \pm 0.2$ & $2.0 \pm 0.3$ &  $0.74 \pm 0.06$ & $0.9 ^{+0.9}_{-0.7} $& 25.9 & $9.0 \pm 0.1$ &  $20 \pm 5$& $20 \pm 5$ & 19.5 & $7.9 \pm 0.1$ &  $9.45 \pm 0.07$& $9.57 \pm 0.07$ & 0.92 (277) \\ 
1RXS~J1718  & $2.1 \pm 0.1$ & $1.5 \pm 0.1$ &  $0.39 \pm 0.03$ &$3.4^{+0.6}_{-0.9}$ & 49.1 & $4.02 \pm 0.05$ &  $9.4 \pm 0.7$& $9.1 \pm 0.4$ & 11.2 & $3.35 \pm 0.05$ &  $3.81 \pm 0.04$& $3.69 \pm 0.02 $ & 1.03 (275) \\
1RXH~J1735 & $1.7 \pm 0.1$ & $1.4 \pm 0.2$  & $0.38 \pm 0.03$ & $3.2^{+0.8}_{-1.5}$ & 53.8 &  $2.66 \pm 0.07$ & $5.9 \pm 0.7$ & $6.4 \pm 0.7$ & 11.6 &  $2.11 \pm 0.07$ & $2.33 \pm 0.06$ & $2.52 \pm 0.06$  & 0.97 (184) \\

\hline

\end{tabular}
\label{tab:spec}
\begin{tablenotes}
\item[]Note. -- Quoted errors represent 90\% confidence levels. The fifth and ninth column reflects the fractional contribution of the thermal component to the total unabsorbed 0.5--10 keV and 2--10 keV flux respectively. $F_{\mathrm{X, abs}}$ and $F_{\mathrm{X, unabs}}$ represent the absorbed and unabsorbed fluxes, respectively. The luminosity $L_{\mathrm{X}}$, source radius ($R_{\mathrm{bb}}$) and apparent inner disk radius ($R_{\mathrm{in}}$) were calculated adopting distances of 9.2, 9, and 9.5 kpc for \ax, \rxs\ and \rxh, respectively. Also we assumed a disk angle $\theta=0 $ to calculate $R_{\mathrm{in}}$.
\end{tablenotes}
\end{threeparttable}
\end{sideways}
\end{table*}


%
%
%
%
%
%
%


\section{Discussion}\label{sec:discussion}
We have presented the spectral analysis of our high-quality \xmm\ data of the persistent neutron star LMXBs \ax, \rxs, and \rxh. Their 2--10 keV X-ray luminosities are $L_{\mathrm{X}} \simeq (2-10) \times 10^{34}~\lum$, which implies that these neutron stars are accreting at $\simeq 0.01\%-0.05\%$ of the Eddington rate \citep[assuming an Eddington luminosity of $L_{\mathrm{EDD}} = 3.8 \times10^{38}~\lum$;][]{kuulkers2003}.\\

Previous \swift\ and \chan\ observations of \rxs\ could be satisfactorily modeled with an absorbed power-law model ($\Gamma \simeq 2.1-2.4$; \citealt[][]{zand2005,zand2009}). Fitting our XMM-Newton data of the source with a single power-law model gives as similar photon index. However, due to the quality of our data it has now been possible to determine that the spectrum can be better described by a two component model, consisting of a power-law plus a soft component. This soft component could be satisfactory be modeled with either a black-body or a disk black body model. Such a two component model results in a significantly harder photon index ($\Gamma\sim1.5$) compared to the single power-law model. The temperature of the soft component is $\sim0.3-0.4$~keV. 

We obtained similar results for \rxh . The values obtained with the absorbed power--law model are in agreement with the ones reported by \citet[][]{degenaar2010_rxh} using \swift\ data (they obtained a photon index of $\Gamma\sim2.3$ compared to our 2.5). However, the \xmm\ spectral analysis shows the presence of a thermal component with a temperature $\sim0.3-0.4$~keV and consequently, the photon index has (similarly to \rxs) a lower value ($\Gamma\sim1.4$). \ax\ is the only one of the three sources that is well described by only a simple absorbed power--law model and the addition of a soft spectral component did not improve the fit. This is consistent with the results obtained by \citet{degenaar2012_asca} and \citet{DelSanto2008} who analyzed \swift\ observations obtained from this source. \ax\ has the highest column density of the three sources ($N_{H}\sim2.9\times10^{22}\nh$) but we determined that this high column density could not explain why we do not see a soft component in its spectrum with similar strength as what we observed for the other two sources (see below). Another difference is that \ax\ is the  most luminous of the three sources ($L_{\mathrm{X}}\sim10^{35}~\lum$ for \ax\ compared with the $L_{\mathrm{X}}\sim3-4\times10^{34}~\lum$ for the other two sources; 2--10~keV).\\

One question is why the soft component is detected in some sources and not in others. The data quality of our three sources are similar; in fact the \ax\ data are slightly better. Hence the spectral differences are not due to differences in the quality of the data. One possibility is the difference in the absorption by the interstellar medium along the line of sight towards the sources. We have carried out simulations in order to quantify the effect of the high $N_{H}$ on our ability to detect a soft component in the spectra. Using \textsc{XSpec} we simulated a variety of spectra using the RMFs and ARFs for the MOS cameras, an exposure time of 30~ks and the parameters obtained for \rxs\ and \rxh\ but with an $N_{H}$ up to $3\times10^{22}\nh$. We found that despite that its contribution to the total absorbed flux is lower, the thermal component is still clearly detectable (albeit at lower significance) if its contribution to the total unabsorbed flux is similar as what has been observed for \rxs\ and \rxh\ ($\sim40\%-50\%$ in the 0.5--10~keV energy range). Therefore, the reason for the spectral differences cannot be explained by differences in absorption columns alone and must be intrinsic to the sources.\\ 

We have searched the literature for information about what is known about the spectral properties of other persistent very-faint X-ray binaries. The NS LMXB \xmmsource\ is such  source, although strictly speaking it is a quasi-persistent LMXB because before 2001 it could not be detected \citep[but is has been detected every since;][]{sidoli2003,delsanto2007,degenaar2011_burstxmmsource}. Its 2--10~keV X-ray luminosity is $L_{\mathrm{X}} \sim 5 \times 10^{34}~\lum$ \citep[][]{delsanto2007}. Its \xmm\ spectra were well described with an absorbed power--law model, with a photon index of $\Gamma\sim2$ but its column density is very high with a value of $N_{H}\sim9\times10^{22}\nh$ which could easily absorb the flux from even a rather strong thermal component. We note that, as discussed above, this cannot be the reason why we do not see a thermal component in \ax\ because the column density of that source is significantly lower than what is observed for \xmmsource. Therefore the absence of a soft thermal component in \ax\ is likely to be intrinsic for this source. The unclassified persistent source \ascasource\ reaches similar luminosities and its spectrum can be described with a similar spectral model. However, this source is also highly absorbed ($N_{H}\sim8\times10^{22}\nh$) making detection of a soft thermal component difficult. Its spectrum is relatively soft \citep[$\Gamma\sim2.3$;][]{degenaar2012_asca}, similarly to \ax.    

In addition, there are several LMXB transients which never become very bright and only reach peak X-ray luminosities similarly to the very-faint persistent sources.  E.g., the best fit model for the \xmm\ spectrum of the very faint LMXB \xte, which most likely harbours a neutron star, was modeled with a blackbody component with a temperature of $\sim0.3$~keV combined with a power--law with a photon index of  $\Gamma\sim1.7$. The source has a rather low column density, $N_{H}\sim3\times10^{21}\nh$ \citep[][]{armaspadilla2011} and the addition of a soft thermal component to the fit model is statistically required, which contributes $\sim30\%$ to the 0.5--10 keV source flux. The 2--10~keV luminosity of the source during this \xmm\ observation was $1.3\times10^{34}~\lum$ which is very similar to that of \rxs\ and \rxh. Similarly, the transient neutron star LMXB Swift~J185003.2--005627 displayed a 0.5--10~keV luminosity of $\sim3\times10^{35}~\lum$ in its 2011 outburst. The best description for the \swift\ spectra was also with a two component model with similar parameter values, and the thermal component contributed $\sim45\%-65\%$ to the total 0.5--10 keV source unabsorbed flux \citep{degenaar2012_swiftbursters}. \\

From the above it is clear that our understanding of the X-ray spectra of very faint X-ray binaries is severely limited and hampered by the lack of sources in combination with the sometimes large absorption to the known sources. It seems that at the lowest luminosities (a few times $10^{34}~\lum$) that definitely a soft component (with a low kT of $\sim0.3-0.4$~keV) is needed and that the power--law component is rather hard (photon index of 1.4--1.6). However, at luminosities of $10^{35}~\lum$ or higher, the spectra either do not require a soft component at similar strength as seen for the fainter sources (e.g. in \ax) or a rather strong soft component could be clearly detected as well (e.g., Swift~J185003.2--005627; albeit with relatively high temperatures of 0.7~keV). The nature of this thermal component is not clear. However, considering that most of the sources harbour neutron stars accretors, a plausible origin for this component is the neutron star surface or perhaps the boundary layer (if present). Low-level accretion onto the surface of a neutron star can indeed produce a black-body like spectrum  \citep{Zampieri1995}. Still, a disc origin cannot be ruled out since it was also possible to fit the data with a multicolor disc blackbody. Moreover, it is not clear that the soft component which can be detected in the different sources is always coming from the same region: it might be possible that in some sources we see the surface/boundary layer of the neutron star but in others we see emission from the disk.\\



Despite that a soft component is not needed in the spectra of all known
sources that have $L_{X}\sim10^{35}~\lum$, the spectra of those sources are still
remarkable soft compared to what is typically seen for sources at higher $L_{X}$
(i.e. $10^{36}~\lum$, with usually $\Gamma\la2$). It is unclear if all the softening can be explained by the introduction of a thermal soft component (e.g., as seen for Swift~J185003.2--005627) since this is not required by the data of \ax\ and will only contribute at most 10--20\% to the 0.5--10~keV flux (see Table 1). Therefore, quite likely in a number of sources the soft spectra at $\sim10^{35}~\lum$ might be intrinsic to the power--law component. This is similar to
what has been seen for bright transients which decay below a
luminosity of $10^{36}~\lum$ as well as for several transient VFXBs \citep{armaspadilla2011, armaspadilla2013}: their spectra become softer as their X-ray luminosities decrease. The fact that at the same low X-ray luminosities similar soft X-ray spectra are seen in persistent sources and in transient sources might suggest that both types of systems have similar accretion geometries in the luminosity regime $10^{35-36}~\lum$. Such softening of the power--law component is expected in ADAF or RIAF like accretion flows \citep[see][for a discussion]{armaspadilla2013}. \\

What happens below a luminosity of $10^{35}~\lum$ is not clear. The presence (in two of our sources) of a soft thermal component at a luminosity close to $10^{34}~\lum$ causes the spectra of those sources to be as soft as observed. In fact, their power-law component seem to be rather hard with photon indices of $\sim$1.5. It is unclear if this power--law component is related to the one seen at higher luminosities (which would indicate that it becomes harder again at the lowest observed luminosities if indeed the softening from $10^{36}~\lum$ down to $10^{35}~\lum$ can be ascribed to changes in the power-law component) or that it is due to a different phenomena (in which case the process which produces the power-law component seen at $>10^{35}~\lum$ must have disappeared and must be replaced by a different process which also produces a power-law shape spectrum). Therefore, currently it is unclear how the lowest luminosity sources fit in the overall picture of accreting X-ray binaries. More sources in this luminosity range have to be studied using very sensitive satellites like XMM-Newton. In particular those sources which move (frequently) through the luminosity range $10^{34-35}~\lum$, such as the transient systems or highly variable systems.\\

The spectral characteristics of \rxs\ and \rxh\ are similar to some of the transient neutron star LMXBs in their quiescence state, especially those who have $L_{X}$ between $10^{33}$ and $10^{34}~\lum$. Also for those sources, their X-ray spectra can be described using a two-component model consisting of a thermal part (fitted usually with a black-body model or a neutron-star atmosphere model) and a non-thermal part (usually fitted with a power-law model). Remarkably the fit parameters obtained using such a model are very similar to what we have obtained for \rxs\ and \rxh: a black body temperature of 0.2--0.3~keV and a photon index of $\sim1.5$ (albeit that this is often not well constraint due to the low quality of the data) \citep[e.g.][]{Rutledge1999,Rutledge2002,Wijnands2002,Wijnands2004,Jonker2003,Cackett2011,degenaar2012_EXO}. In this respect it is interesting to note that when the thermal component in  quiescent systems are fitted with a black--body spectrum, small emitting radii are obtained (too small to be consistent with the expected radii of neutron stars), similarly to what we observe for \rxs\ and \rxh\ (see Table \ref{tab:spec}). To explain those small radii in the quiescent systems, it has been argued that a black--body model is not the correct model to use to fit the thermal component, but instead a neutron--star atmosphere model would be more appropriate \citep[e.g,][]{Brown1998}. Using such models, more realistic neutron star radii are indeed obtained \citep[e.g,][]{Rutledge1999, Rutledge2000}. Applying such models to our targets also result in larger inferred radii ($\sim$10~km) but likely such models cannot be applied to our targets because they are still accreting. More appropriate models \citep[like the one of ][]{Zampieri1995, Soria2011} have to be used for our targets, but it goes beyond the scope of the paper to discuss those models in detail.\\

The interpretation of the non-thermal component remains unclear \citep[see][and references there in]{campana2009,degenaar2012_EXO}. The similarity between these quiescent systems and our persistent accreting sources at a few times $10^{34}~\lum$ could point that accretion might still be going on in quiescence for at least some of those sources which have quiescent luminosities of $10^{33}-10^{34}~\lum$. For systems which are fainter, this cannot be concluded because often either the thermal component is not present and only the power-law component (which often appears to be softer with photon index of $\sim2$; albeit again with large error bars) or only the soft component is seen with hardly any or no detectable contribution by a power--law component. In those latter systems we could indeed observe the cooling of the neutron star. The physical process behind the fully non-thermally dominated quiescent spectra of a growing fraction of the quiescent systems is not understood.

\section*{Acknowledgements} 
M.A.P. acknowledges the hospitality of the University of Michigan, where part of this work was carried out. We thank Alesandro Patruno and Lucy Heil for their useful comments on a previous version of this paper.
This work made use of data from the \xmm\ public data archive. ND is supported by NASA through Hubble Postdoctoral Fellowship grant number HST-HF-51287.01-A from the Space Telescope Science Institute (STScI). RW and MAP are supported by an European Research Counsil starting grant awarded to RW. 

\bibliographystyle{mn2eOK}
\bibliography{VF_persistent}

\label{lastpage}
\end{document}